\pdfoutput=1
\documentclass[12pt,a4paper]{article}

\usepackage{ifthen} 
\newboolean{pdflatex}
\setboolean{pdflatex}{true} 

\newboolean{articletitles}
\setboolean{articletitles}{true} 

\newboolean{uprightparticles}
\setboolean{uprightparticles}{false} 

\newboolean{inbibliography}
\setboolean{inbibliography}{false} 

\def\paperauthors{LHCb collaboration} 
\def\paperasciititle{Observation of} 
\def\papertitle{Observation of Cabibbo-suppressed two-body hadronic decays and precision mass measurement of the \Omegac baryon} 
\def\paperkeywords{{High Energy Physics}, {LHCb}} 
\def\papercopyright{\the\year\ CERN for the benefit of the LHCb collaboration} 
\def\paperlicence{CC BY 4.0 licence}
\def\paperlicenceurl{https://creativecommons.org/licenses/by/4.0/}

\usepackage[top=1in, bottom=1.25in, left=1in, right=1in]{geometry}

\usepackage{microtype}
\usepackage{lineno}  
\usepackage{xspace} 
\usepackage{caption} 

\usepackage{graphicx}  
\usepackage{color}
\usepackage{colortbl}
\graphicspath{{./figs/}} 

\usepackage{amsmath} 
\usepackage{amssymb}
\usepackage{amsfonts}
\usepackage{upgreek} 

\newcommand*\patchAmsMathEnvironmentForLineno[1]{%
\expandafter\let\csname old#1\expandafter\endcsname\csname #1\endcsname
\expandafter\let\csname oldend#1\expandafter\endcsname\csname
end#1\endcsname
 \renewenvironment{#1}%
   {\linenomath\csname old#1\endcsname}%
   {\csname oldend#1\endcsname\endlinenomath}%
}
\newcommand*\patchBothAmsMathEnvironmentsForLineno[1]{%
  \patchAmsMathEnvironmentForLineno{#1}%
  \patchAmsMathEnvironmentForLineno{#1*}%
}
\AtBeginDocument{%
\patchBothAmsMathEnvironmentsForLineno{equation}%
\patchBothAmsMathEnvironmentsForLineno{align}%
\patchBothAmsMathEnvironmentsForLineno{flalign}%
\patchBothAmsMathEnvironmentsForLineno{alignat}%
\patchBothAmsMathEnvironmentsForLineno{gather}%
\patchBothAmsMathEnvironmentsForLineno{multline}%
\patchBothAmsMathEnvironmentsForLineno{eqnarray}%
}

\usepackage{hyperxmp}

\usepackage[pdftex,
            pdfauthor={\paperauthors},
            pdftitle={\paperasciititle},
            pdfkeywords={\paperkeywords},
            pdfcopyright={Copyright (C) \papercopyright},
            pdflicenseurl={\paperlicenceurl}]{hyperref}

\usepackage[colorinlistoftodos,textsize=scriptsize]{todonotes}

\usepackage[bottom,flushmargin,hang,multiple]{footmisc}

\usepackage[all]{hypcap} 

\usepackage{xspace} 
\usepackage{upgreek}

\def\lhcb   {\mbox{LHCb}\xspace}

\def\MagUp {\mbox{\em Mag\kern -0.05em Up}\xspace}

\ifthenelse{\boolean{uprightparticles}}%
{

 \def\Ppi         {\ensuremath{\uppi}\xspace}

 \def\PDelta      {\ensuremath{\Delta}\xspace}                 
 \def\PXi         {\ensuremath{\Xi}\xspace}                 
 \def\PLambda     {\ensuremath{\Lambda}\xspace}                 
 \def\PSigma      {\ensuremath{\Sigma}\xspace}                 
 \def\POmega      {\ensuremath{\Omega}\xspace}                 
 \def\PUpsilon    {\ensuremath{\Upsilon}\xspace}

 \def\PB      {\ensuremath{\mathrm{B}}\xspace}                 
                  
 \def\PD      {\ensuremath{\mathrm{D}}\xspace}

 \def\PK      {\ensuremath{\mathrm{K}}\xspace}

 \def\PW      {\ensuremath{\mathrm{W}}\xspace}

 \def\Pc      {\ensuremath{\mathrm{c}}\xspace}

 \def\Pi      {\ensuremath{\mathrm{i}}\xspace}

 \def\Pp      {\ensuremath{\mathrm{p}}\xspace}

 \def\Ps      {\ensuremath{\mathrm{s}}\xspace}

 \def\thebaroffset{0.0em}
}
{

 \def\Ppi         {\ensuremath{\pi}\xspace}

 \mathchardef\PDelta="7101
 \mathchardef\PXi="7104
 \mathchardef\PLambda="7103
 \mathchardef\PSigma="7106
 \mathchardef\POmega="710A
 \mathchardef\PUpsilon="7107
                  
 \def\PB      {\ensuremath{B}\xspace}                 
                  
 \def\PD      {\ensuremath{D}\xspace}

 \def\PK      {\ensuremath{K}\xspace}

 \def\PW      {\ensuremath{W}\xspace}

 \def\Pc      {\ensuremath{c}\xspace}

 \def\Pi      {\ensuremath{i}\xspace}

 \def\Pp      {\ensuremath{p}\xspace}

 \def\Ps      {\ensuremath{s}\xspace}

 \def\thebaroffset{0.18em}
}
\newcommand{\offsetoverline}[2][\thebaroffset]{\kern #1\overline{\kern -#1 #2}}%

\makeatletter
\ifcase \@ptsize \relax
  \newcommand{\miniscule}{\@setfontsize\miniscule{4}{5}}
\or
  \newcommand{\miniscule}{\@setfontsize\miniscule{5}{6}}
\or
  \newcommand{\miniscule}{\@setfontsize\miniscule{5}{6}}
\fi
\makeatother

\DeclareRobustCommand{\optbar}[1]{\shortstack{{\miniscule (\rule[.5ex]{1.25em}{.18mm})}
  \\ [-.7ex] $#1$}}

\def\W      {{\ensuremath{\PW}}\xspace}

\def\squark    {{\ensuremath{\Ps}}\xspace}

\def\cquark    {{\ensuremath{\Pc}}\xspace}

\def\pion   {{\ensuremath{\Ppi}}\xspace}

\def\pip    {{\ensuremath{\pion^+}}\xspace}
\def\pim    {{\ensuremath{\pion^-}}\xspace}

\def\kaon    {{\ensuremath{\PK}}\xspace}

\def\KorKbar {\kern \thebaroffset\optbar{\kern -\thebaroffset \PK}{}\xspace}

\def\Kp      {{\ensuremath{\kaon^+}}\xspace}
\def\Km      {{\ensuremath{\kaon^-}}\xspace}

\def\D       {{\ensuremath{\PD}}\xspace}

\def\DorDbar {\kern \thebaroffset\optbar{\kern -\thebaroffset \PD}\xspace}

\def\Dp      {{\ensuremath{\D^+}}\xspace}
\def\Dm      {{\ensuremath{\D^-}}\xspace}

\def\DpDm    {\ensuremath{\Dp {\kern -0.16em \Dm}}\xspace}

\def\B       {{\ensuremath{\PB}}\xspace}

\def\BorBbar {\kern \thebaroffset\optbar{\kern -\thebaroffset \PB}\xspace}

\def\Bd      {{\ensuremath{\B^0}}\xspace}

\def\BdorBdbar {\kern \thebaroffset\optbar{\kern -\thebaroffset \Bd}\xspace}

\def\Bs      {{\ensuremath{\B^0_\squark}}\xspace}

\def\BsorBsbar {\kern \thebaroffset\optbar{\kern -\thebaroffset \Bs}\xspace}

\def\Y#1S{\ensuremath{\PUpsilon{(#1S)}}\xspace}

\def\proton      {{\ensuremath{\Pp}}\xspace}

\def\Lz          {{\ensuremath{\PLambda}}\xspace}

\def\LorLbar     {\kern \thebaroffset\optbar{\kern -\thebaroffset \PLambda}\xspace}

\def\Sigmares    {{\ensuremath{\PSigma}}\xspace}

\def\Xires       {{\ensuremath{\PXi}}\xspace}

\def\Xim         {{\ensuremath{\Xires^-}}\xspace}

\def\Omegares    {{\ensuremath{\POmega}}\xspace}

\def\Omegam      {{\ensuremath{\Omegares^-}}\xspace}

\def\Lcp         {{\ensuremath{\Lz^+_\cquark}}\xspace}
\def\Sigmac      {{\ensuremath{\Sigmares_\cquark}}\xspace}
\def\Xic         {{\ensuremath{\Xires_\cquark}}\xspace}

\def\Omegac      {{\ensuremath{\Omegares^0_\cquark}}\xspace}

\newcommand{\decay}[2]{\ensuremath{#1\!\to #2}\xspace} 

\def\to                 {\ensuremath{\rightarrow}\xspace}

\def\OcToOmpi     {\decay{\Omegac}{\Omegam\pip}}
\def\OcToOmK      {\decay{\Omegac}{\Omegam\Kp}}
\def\OcToXipi     {\decay{\Omegac}{\Xim\pip}}
\def\OmToLzK      {\decay{\Omegam}{\Lz\Km}}
\def\XiToLzpi      {\decay{\Xim}{\Lz\pim}}

\def\AT#1     {\ensuremath{A_{\mathrm{T}}^{#1}}\xspace}           

\def\C#1      {\ensuremath{\mathcal{C}_{#1}}\xspace}                       
\def\Cp#1     {\ensuremath{\mathcal{C}_{#1}^{'}}\xspace}                    
\def\Ceff#1   {\ensuremath{\mathcal{C}_{#1}^{\mathrm{(eff)}}}\xspace}        
\def\Cpeff#1  {\ensuremath{\mathcal{C}_{#1}^{'\mathrm{(eff)}}}\xspace}       
\def\Ope#1    {\ensuremath{\mathcal{O}_{#1}}\xspace}                       
\def\Opep#1   {\ensuremath{\mathcal{O}_{#1}^{'}}\xspace}                    

\newcommand{\aunit}[1]{\ensuremath{\text{\,#1}}}       

\newcommand{\tev}{\aunit{Te\kern -0.1em V}\xspace}
\newcommand{\gev}{\aunit{Ge\kern -0.1em V}\xspace}
\newcommand{\mev}{\aunit{Me\kern -0.1em V}\xspace}
\newcommand{\kev}{\aunit{ke\kern -0.1em V}\xspace}
\newcommand{\ev}{\aunit{e\kern -0.1em V}\xspace}
 
\newcommand{\mevc}{\ensuremath{\aunit{Me\kern -0.1em V\!/}c}\xspace}
\newcommand{\gevc}{\ensuremath{\aunit{Ge\kern -0.1em V\!/}c}\xspace}
\newcommand{\mevcc}{\ensuremath{\aunit{Me\kern -0.1em V\!/}c^2}\xspace}
\newcommand{\gevcc}{\ensuremath{\aunit{Ge\kern -0.1em V\!/}c^2}\xspace}

\def\fb   {\ensuremath{\aunit{fb}}\xspace}
\def\invfb   {\ensuremath{\fb^{-1}}\xspace}

\def\ps   {\ensuremath{\aunit{ps}}\xspace}

\newcommand{\stat}{\aunit{(stat)}\xspace}
\newcommand{\syst}{\aunit{(syst)}\xspace}
\newcommand{\ext}{\aunit{(ext)}\xspace}

\newcommand{\chisq}{\ensuremath{\chi^2}\xspace}

\def\gsim{{~\raise.15em\hbox{$>$}\kern-.85em
          \lower.35em\hbox{$\sim$}~}\xspace}
\def\lsim{{~\raise.15em\hbox{$<$}\kern-.85em
          \lower.35em\hbox{$\sim$}~}\xspace}

\def\pt         {\ensuremath{p_{\mathrm{T}}}\xspace}

\def\evtgen     {\mbox{\textsc{EvtGen}}\xspace}

\def\geant      {\mbox{\textsc{Geant4}}\xspace}

\def\photos     {\mbox{\textsc{Photos}}\xspace}

\def\pythia     {\mbox{\textsc{Pythia}}\xspace}

\def\tell1  {TELL1\xspace}
\def\ukl1   {UKL1\xspace}

\newcommand{\lhcborcid}[1]{\href{https://orcid.org/#1}{\hspace*{0.1em}\raisebox{-0.45ex}{\includegraphics[width=1em]{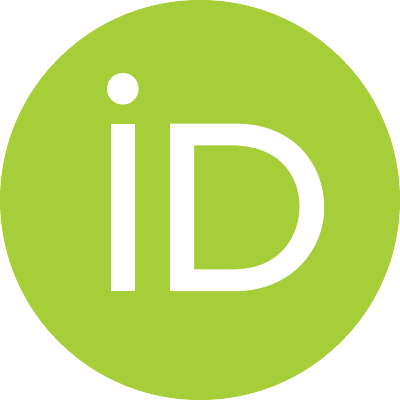}}}}

\usepackage{cite} 
\usepackage{mciteplus}

\usepackage{longtable} 

\begin{document}

\renewcommand{\thefootnote}{\fnsymbol{footnote}}
\setcounter{footnote}{1}


\begin{titlepage}
\pagenumbering{roman}

\vspace*{-1.5cm}
\centerline{\large EUROPEAN ORGANIZATION FOR NUCLEAR RESEARCH (CERN)}
\vspace*{1.5cm}
\noindent
\begin{tabular*}{\linewidth}{lc@{\extracolsep{\fill}}r@{\extracolsep{0pt}}}
\ifthenelse{\boolean{pdflatex}}
{\vspace*{-1.5cm}\mbox{\!\!\!\includegraphics[width=.14\textwidth]{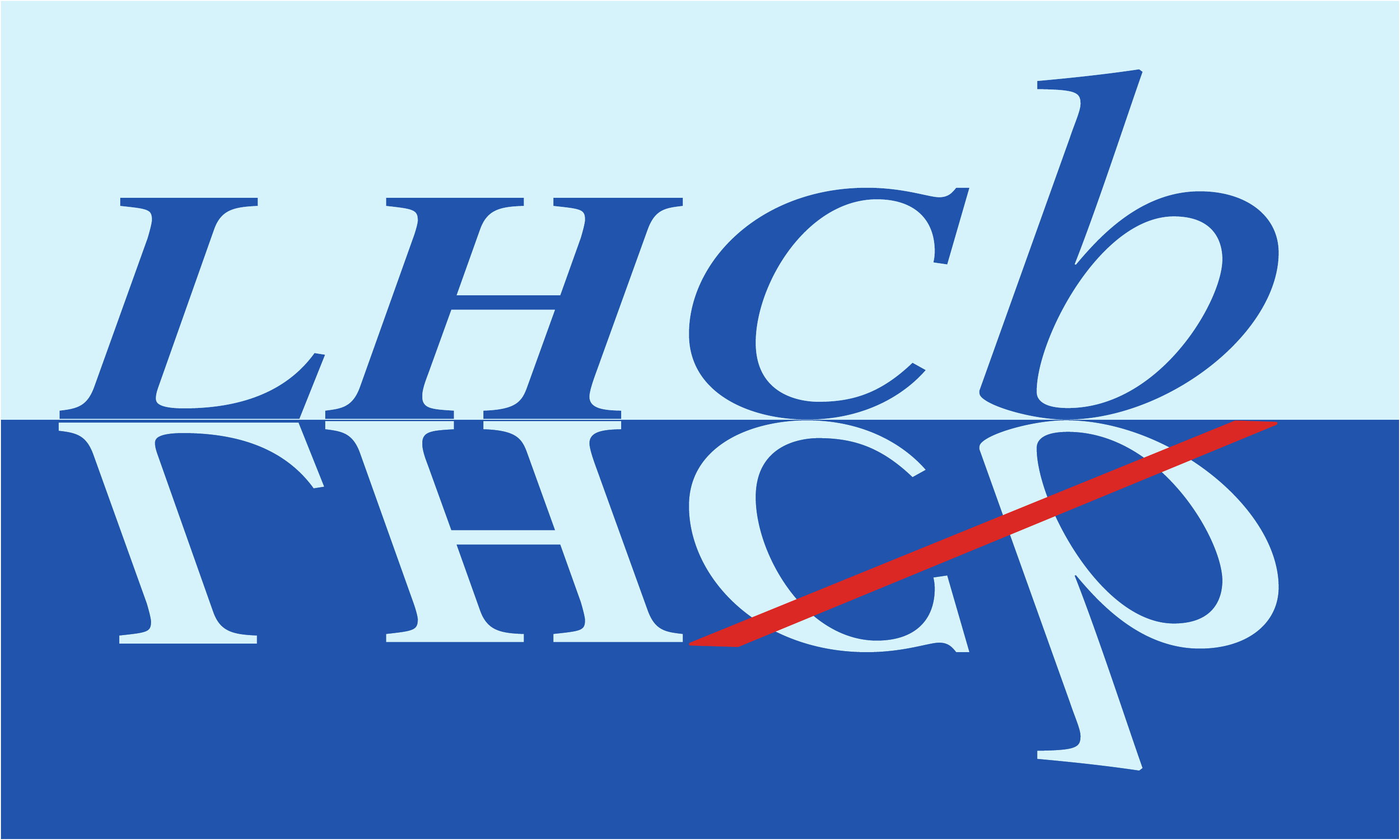}} & &}%
{\vspace*{-1.2cm}\mbox{\!\!\!\includegraphics[width=.12\textwidth]{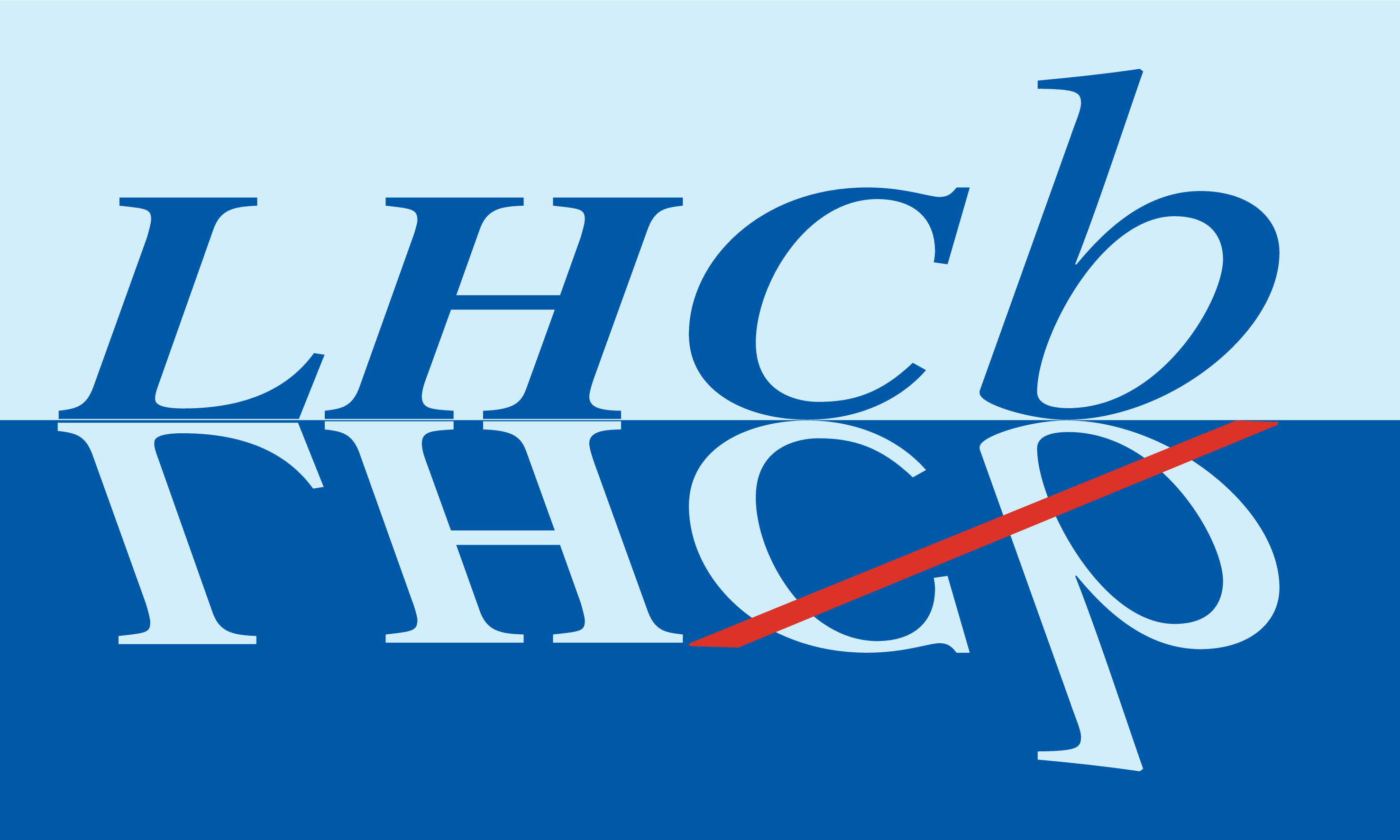}} & &}%
\\
 & & CERN-EP-2023-155 \\  
 & & LHCb-PAPER-2023-011 \\  
 & & February 26, 2024
\end{tabular*}

\vspace*{2.5cm}

{\normalfont\bfseries\boldmath\huge
\begin{center}
  \papertitle 
\end{center}
}

\vspace*{1.0cm}

\begin{center}
\paperauthors\footnote{Authors are listed at the end of this Letter.}
\end{center}

\vspace{\fill}

\begin{abstract}
 \noindent
The first observation of the singly Cabibbo-suppressed \OcToOmK and \OcToXipi decays is reported, 
using proton-proton collision data at a center-of-mass energy of 13\tev, corresponding to an integrated luminosity of 5.4\invfb, collected with the LHCb detector between 2016 and 2018.
The branching fraction ratios are measured to be
\begin{eqnarray}
\frac{\mathcal{B}(\OcToOmK)}{\mathcal{B}(\OcToOmpi)}
&=&[6.08\pm0.51\stat\pm 0.40\syst]\%\text{,}\nonumber\\
\frac{\mathcal{B}(\OcToXipi)}{\mathcal{B}(\OcToOmpi)}
&=&[15.81\pm0.87\stat\pm0.44\syst\pm0.16\ext]\%\text{.}\nonumber
\end{eqnarray}
In addition, using the $\OcToOmpi$ decay channel, the \Omegac baryon mass is measured to be
\begin{eqnarray}
M(\Omegac)&=&2695.28\pm0.07\stat\pm0.27\syst\pm0.30\ext\mev\text{,}\nonumber
\end{eqnarray}
improving the precision of the previous world average by a factor of 4.

\end{abstract}

\vspace*{2.0cm}

\begin{center}
  Published in
  Phys.~Rev.~Lett.~132 (2024) 081802 
\end{center}

\vspace{\fill}

{\footnotesize 
\centerline{\copyright~\papercopyright. \href{\paperlicenceurl}{\paperlicence}.}}
\vspace*{2mm}

\end{titlepage}


\newpage
\setcounter{page}{2}
\mbox{~}
%
%
%
%


\renewcommand{\thefootnote}{\arabic{footnote}}
\setcounter{footnote}{0}

\cleardoublepage


\pagestyle{plain} 
\setcounter{page}{1}
\pagenumbering{arabic}



Charmed baryons have been widely studied in the last decades, providing an excellent sector to scrutinize the dynamics of light quarks bound to a heavy quark~\cite{cite:Cheng_CBP}.
The \Lcp, \Sigmac, and \Xic baryons, as well as their excited states, have been studied experimentally more extensively than the \Omegac baryon.
Some of the current limitations on experimental knowledge of the \Omegac baryon system include the significant uncertainty in the mass determination~\cite{PDG2022,cite:omc0_mass_E687,cite:omc0_mass_cleo,cite:omc0_mass_belle}, recent lifetime results much larger than the previous world-average~\cite{LHCb-PAPER-2018-028,LHCb-PAPER-2021-021,Belle-II:2022plj,Cheng:2021vca}, and the fact that only a few decay modes have been observed~\cite{PDG2022}.
Hence, it is important to extend the knowledge of the \Omegac decay parameters with additional experimental inputs.

Decays of the \Omegac baryon proceed via weak interactions and have been a subject of significant theoretical interest. The decay amplitudes from nonfactorizable contributions, such as internal \W-emission and \W-exchange transitions, play a crucial role in these decays, and various methods have been developed to calculate such contributions~\cite{cite:Cheng_CBP}.
Branching fractions (BFs) of hadronic \Omegac baryon decays are predicted by different theoretical models, including the pole model~\cite{cite:pole_model,cite:Omegac_Omegapi,cite:predict_xuq}, the covariant confined quark model~\cite{cite:cov1,cite:cov2,cite:predict_guts}, current algebra~\cite{cite:predict_xu}, and the light-front quark model~\cite{cite:predict_tsai,cite:predict_zhao}.
Nevertheless, predictions of BFs for both \OcToOmpi and \OcToXipi decays (Charge-conjugate processes are implied throughout this Letter.) have significant discrepancies between the different models~\cite{cite:Omegac_Omegapi,cite:predict_xuq,cite:predict_guts,cite:predict_xu,cite:predict_tsai,cite:predict_zhao}, while there is no theoretical prediction for the \OcToOmK decay.
The \OcToOmK and the \OcToXipi decays occur through external \W-emission or \W-exchange transitions, receiving both factorizable and nonfactorizable contributions.
Conversely, the \OcToOmpi decay is driven only by a factorizable contribution generated by an external \W-emission amplitude.
To this day, no absolute BF of \Omegac decays has been measured. The \OcToOmpi decay is chosen as a normalization channel in all reported studies~\cite{PDG2022}.
Furthermore, no singly Cabibbo-suppressed two-body decay of \Omegac baryons into final states with a \Xim or \Omegam baryon has previously been observed. Only evidence of the \OcToXipi decay has been reported by the Belle Collaboration~\cite{cite:belle_xipi_omk}.
In addition, the latest \Omegac mass measurement reported by Belle is statistically limited by the small sample size of \OcToOmpi decays used for the measurement~\cite{cite:omc0_mass_belle}. This situation affects precision measurements of heavier hadrons that decay to the \Omegac baryon.
Also, it should be noted that the obtained \Omegac masses in different lattice quantum chromodynamics predictions~\cite{cite:LQCD1,cite:LQCD2,cite:LQCD3,cite:LQCD4,cite:LQCD5,cite:LQCD6,cite:LQCD7,cite:LQCD8} are overall significantly smaller than the experimental result, while the predicted masses from other theoretical models~\cite{cite:Other1,cite:Other2,cite:Other3,cite:Other4,cite:Other5} are larger than the experiment result. 
Precise measurements of BFs and the mass of the \Omegac baryon are therefore necessary to test the theoretical models and to improve the understanding of the weak interactions in the charmed baryon sector.

This Letter reports the first observation of \OcToOmK and \OcToXipi decays and a precision measurement of the \Omegac mass. The \Omegam (\Xim) candidates are reconstructed through their decay to the \Lz\Km (\Lz\pim) final state, where the \Lz baryons are reconstructed in the \proton\pim final state.
This measurement is based on a data sample collected in proton-proton (\proton\proton) collisions with the \lhcb detector between 2016 and 2018 at a center-of-mass energy of 13\tev, corresponding to an integrated luminosity of 5.4\invfb.
The normalization channel \OcToOmpi has a large yield and is used to measure the \Omegac mass.


The \lhcb detector is a single-arm forward
spectrometer covering the \mbox{$2<\eta <5$} pseudorapidity range, described in detail in Refs.~\cite{LHCb-DP-2008-001, LHCb-DP-2014-002, LHCb-DP-2014-001, LHCb-DP-2013-003, LHCb-DP-2012-002}.
The online event selection is performed by a trigger~\cite{LHCb-DP-2019-002}, comprising a hardware stage based on information from the calorimeter system, followed by a software stage that applies a full event reconstruction. 
The software trigger relies on identifying \Omegam (\Xim) baryon decays to \Lz\Km (\Lz\pim) combinations, and \Kp/\pip tracks consistent with originating from a \Omegac baryon decay vertex. 

Samples of simulated events are used to optimize selection requirements and estimate the efficiencies of the signal and the normalization channels.
The simulated \proton\proton collisions are generated using
\pythia~\cite{Sjostrand:2007gs} with a specific \lhcb
configuration~\cite{LHCb-PROC-2010-056}. Decays of hadronic particles and interactions with the detector material are described by \evtgen~\cite{Lange:2001uf}, using \photos~\cite{davidson2015photos}, and by the \geant toolkit~\cite{Allison:2006ve, *Agostinelli:2002hh, LHCb-PROC-2011-006}, respectively. 
Simulated samples for signal and normalization channels are generated using a uniform phase-space distribution. 


Good-quality tracks with transverse momentum \mbox{$\pt>100\mev$} and momentum \mbox{$p>1\gev$} are selected to form final-state hadrons. (Natural units with \mbox{$\hbar=c=1$} are used throughout this Letter.)
By using dedicated neural networks, particle identification (PID) is performed using the information from all the subdetector systems~\cite{LHCb-DP-2018-001}.
All final-state hadrons must have PID information consistent with the corresponding particle mass hypothesis.
These hadrons are required to be inconsistent with originating from a primary \proton-\proton collision vertex (PV). This condition is achieved by selecting tracks with a large impact parameter significance \mbox{$\chisq_{\rm IP}$}, defined as the \mbox{$\chisq$} difference of a given PV fit with and without the particle (here, \proton, \Km, or \pim) under consideration.
Given the long lifetimes of the \Lz candidates and since they are decay products of two-stage cascade decays of particles with similarly long lifetimes, the \Lz decay products are reconstructed outside the vertex locator.
Protons and pions originating from a \Lz decay are required to have momentum greater than 3\gev. 
Each \Lz candidate must have a good-quality vertex and an invariant mass within \mbox{$6\mev$} of the known \Lz mass~\cite{PDG2022}.
The associated \Km (\pim) particles originating from \Omegam (\Xim) baryon decays are required to have \mbox{$\chisq_{\rm IP}>16$} to suppress the prompt background produced at the \proton\proton collision point.
Each \Omegam (\Xim) candidate is required to have a transverse momentum greater than 500\mev, a reconstructed decay time greater than 2\ps, a good-quality vertex, and an invariant mass within \mbox{$8\mev$} of the known \Omegam (\Xim) baryon mass~\cite{PDG2022}.

The signal and normalization channels are reconstructed by combining \Omegam(\Xim) and \pip(\Kp) candidates, where the well-identified additional pions or kaons are selected by requiring \mbox{$\chisq_{\rm IP}>4$}.
The \Omegac candidates must have a small \mbox{$\chisq_{\rm IP}$} and a positive decay time with respect to its associated PV, and should form a good-quality decay vertex.
The associated PV is the one for which the \Omegac candidate has the smallest \mbox{$\chisq_{\rm IP}$}.
The \Omegac candidates are also required to have \mbox{$\pt>800\mev$} and an invariant mass within \mbox{$45\mev$} of the known \Omegac mass~\cite{PDG2022}.
A kinematic fit~\cite{Hulsbergen:2005pu} of the decay chain constrains the \Omegac candidate to originate from the associated PV, and the \Omegam/\Xim and \Lz candidates to have their known masses~\cite{PDG2022}.
The four-momenta of all the final-state particles are updated accordingly.

\begin{figure}[tp]
  \begin{center}
    \includegraphics[width=0.325\linewidth]{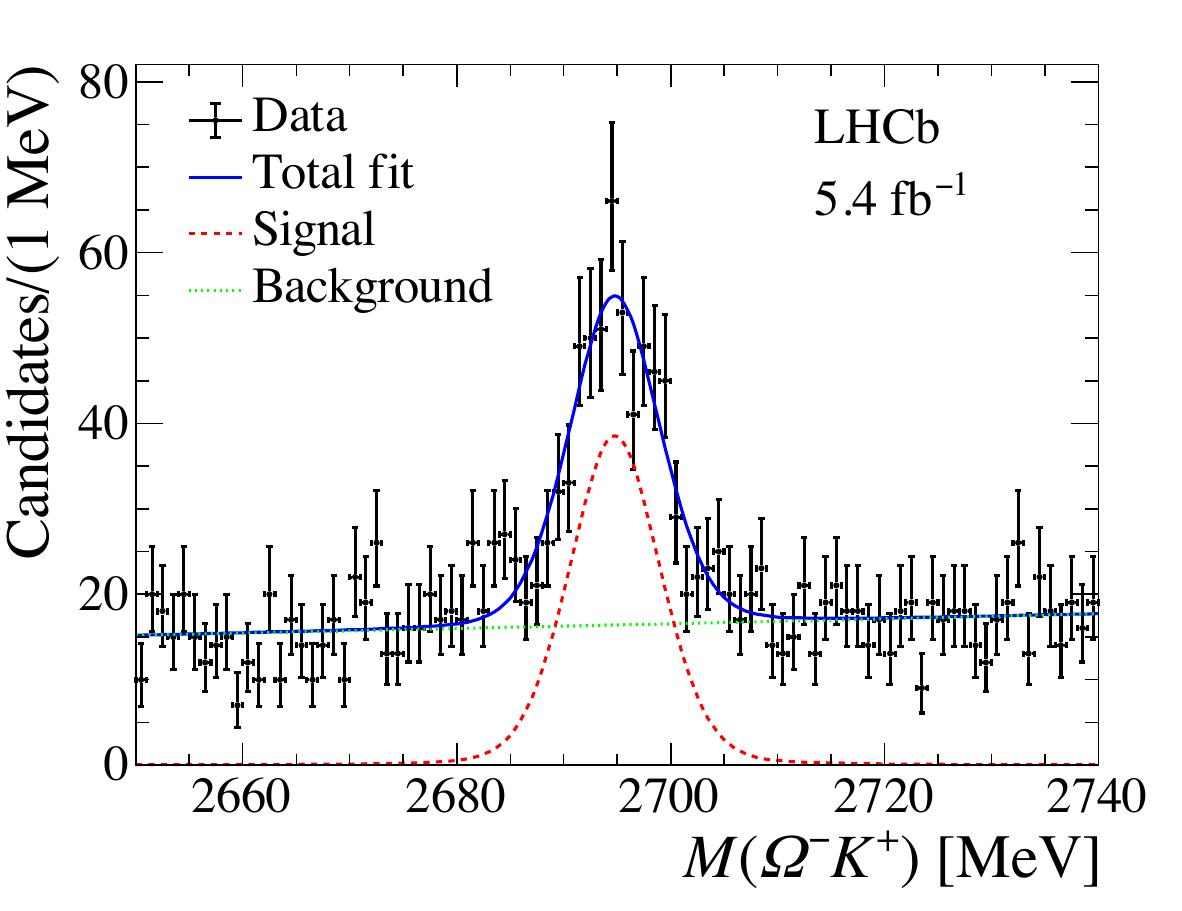}
    \includegraphics[width=0.325\linewidth]{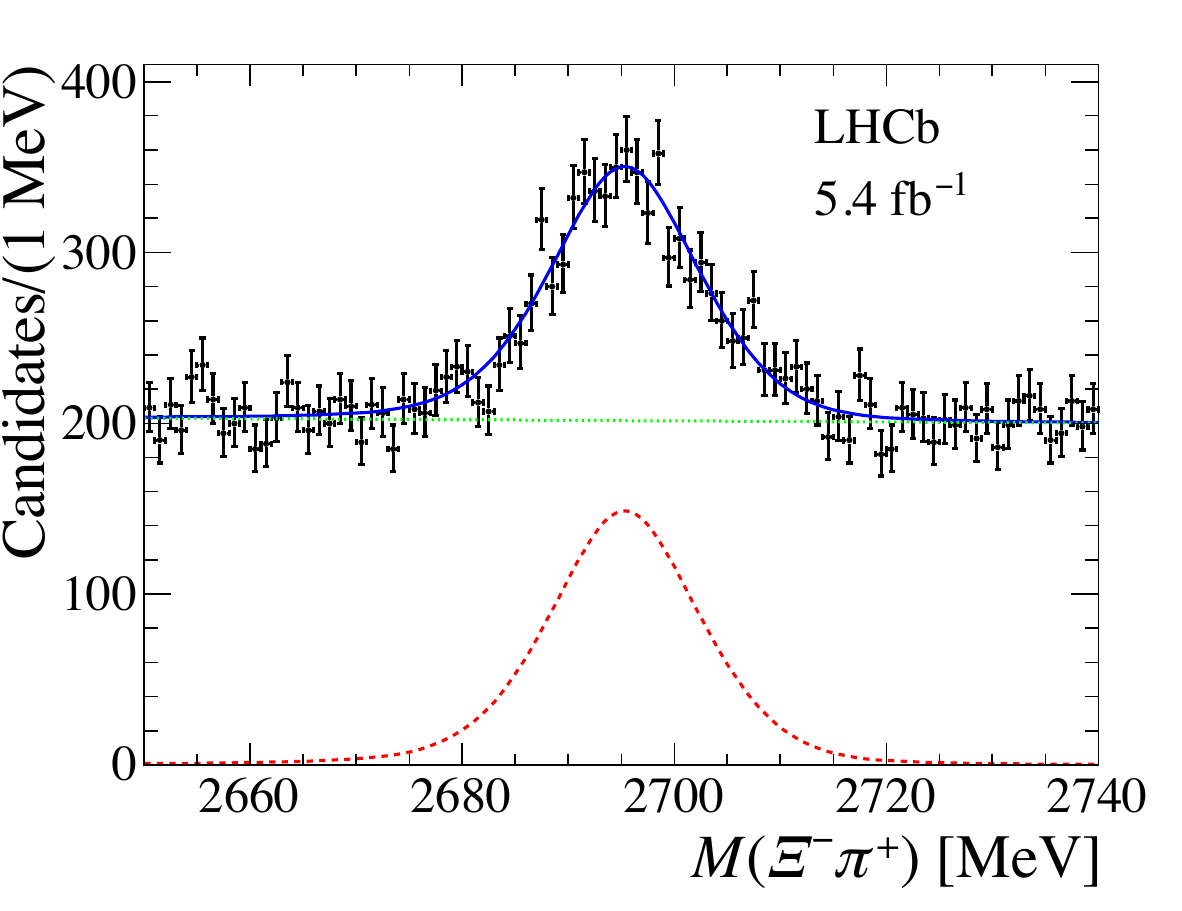}
    \includegraphics[width=0.325\linewidth]{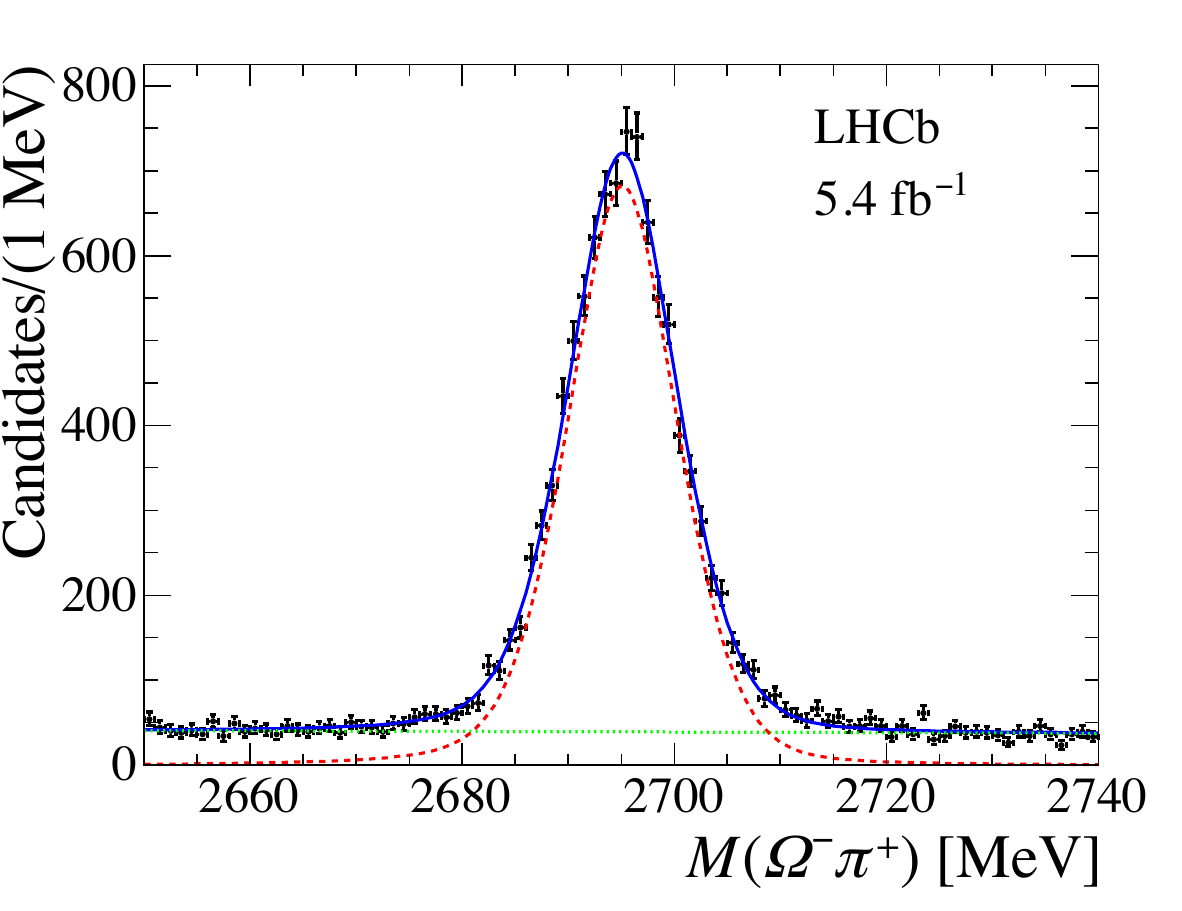}
    \vspace*{-0.5cm}
  \end{center}
  \caption{
    Invariant-mass distributions for (left) \OcToOmK, (middle) \OcToXipi, and (right) \OcToOmpi decays.}
  \label{fig:fit_results}
\end{figure}

After applying the selection criteria, an extended unbinned maximum-likelihood fit is performed to the \Omegam\Kp, \Xim\pip, and \Omegam\pip invariant-mass distributions shown in Fig.~\ref{fig:fit_results}, resulting in signal yields of \mbox{$425\pm35$}, \mbox{$2780\pm150$}, and \mbox{$9330\pm110$}, respectively.
The \Omegac signal shapes are described by the sum of a Gaussian function and a Johnson \mbox{$S_U$} distribution~\cite{cite:johnson} sharing the same mean and width parameters determined from the fit to data (baseline model).
The tail parameters of the Johnson \mbox{$S_U$} function and the fractions for the components are fixed to values obtained from a fit to simulated events.
The background contribution arises only due to random combinations of charged particles in the event. This component is modeled by an exponential function, whose parameters are allowed to vary freely in the fit and to be different between the signal and normalization channels.

From the fit to the \Omegam\pip invariant-mass distribution, the \Omegac baryon mass is measured to be \mbox{$2695.28\pm0.07\mev$}, where the uncertainty is statistical only.
Table~\ref{tab:mass_syst} summarizes the systematic uncertainties on the measurement of the \Omegac baryon mass, which are dominated by the momentum-scale calibration and the uncertainty on the known value of the mass of the \Omegam baryon.
The momentum-scale uncertainty is assessed by shifting the momentum of all charged tracks by \mbox{$\pm0.03\%$}~\cite{LHCb-PAPER-2012-048,LHCb-PAPER-2013-011} in the simulated samples, resulting in a change of 0.27\mev in the \Omegac mass.
In the simulation, the amount of material traversed by a charged particle in the tracking system is known to 10\% accuracy~\cite{LHCb-PAPER-2010-001}.
A systematic uncertainty of 0.03\mev for the energy loss correction due to the uncertainty of the material interaction lengths in the simulation is assigned after scaling by the number of final-state particles~\cite{LHCb-PAPER-2013-011}.
Pseudoexperiments are performed to evaluate the uncertainty due to the choice of the fit model, by generating the \Omegac mass spectrum with the baseline model described above and fitting it with an alternative model. 
The alternative invariant-mass model for the signal consists of a Crystal Ball function~\cite{Skwarnicki:1986xj} combined with a Johnson \mbox{$S_U$} distribution, while the alternative background model is a linear function.
The resulting mass shift of 0.01\mev is assigned as a systematic uncertainty of the invariant-mass fit model.
The total systematic uncertainty, obtained by adding all contributions in quadrature, is determined to be 0.27\mev.
To compute the invariant mass of the \Omegac candidates, the known masses of the \Omegam and \Lz baryons~\cite{PDG2022} are used as constraints in the kinematic fit, and their uncertainties, combining to 0.30\mev, are taken as a systematic uncertainty due to external input. 
\begin{table}[tp]
\setlength{\abovecaptionskip}{0cm}
\setlength{\belowcaptionskip}{-0.3cm}
  \caption{
    Systematic uncertainties for the \Omegac mass measurement.}
\begin{center}\begin{tabular}{lc}
    \hline
    Source & Uncertainty [$\mev$]\\
    \hline
    Momentum scale calibration & $0.27$\\
    Energy loss correction & $0.03$\\
    Fit model & $0.01$\\
    \hline
    Total  & 0.27\\
    \hline
    External input masses  & 0.30\\
    \hline
  \end{tabular}\end{center}
\label{tab:mass_syst}
\end{table}

The BF ratios are calculated as 
\begin{equation}
    \frac{\mathcal{B}(\OcToOmK)}{\mathcal{B}(\OcToOmpi)}=\frac{r_N}{r_{\epsilon}}\text{,}~~~
    \frac{\mathcal{B}(\OcToXipi)}{\mathcal{B}(\OcToOmpi)}=\frac{r'_N}{r'_{\epsilon}}\cdot \frac{\mathcal{B}(\OmToLzK)}{\mathcal{B}(\XiToLzpi)}\text{,}
    \label{eq:BFratio}
\end{equation}
where \mbox{$r^{(\prime)}_N$} is the ratio of yields between the signal and normalization channels, \mbox{$r^{(\prime)}_{\epsilon}$} is the corresponding ratio of total efficiencies, \mbox{$\mathcal{B}(\OmToLzK)=(67.8\pm0.7)\%$} and \mbox{$\mathcal{B}(\XiToLzpi)=(99.887\pm0.035)\%$} are the latest world averages~\cite{PDG2022}.
The total efficiencies include the geometrical acceptance and the reconstruction, trigger and selection efficiencies, which are determined from simulated samples.
Various corrections to the simulated samples are applied to ensure good agreement between data and simulation. The simulated PID variables used as input to the neural network algorithm for each charged track have been calibrated using dedicated high-statistics data samples.
To compute the efficiency, the distributions considered in the event selection are corrected in the simulated samples to match the corresponding signal distributions, where the background is statistically subtracted. Owing to the similarity in the decay topology of the signal and normalization channels, the difference between signal-weighted data and simulation is obtained using the \OcToOmpi sample, which has the largest signal yield. This correction factor is applied to all simulated signal samples.
The overall ratios of efficiencies, \mbox{$r_{\epsilon}$} and \mbox{$r'_{\epsilon}$}, are found to be \mbox{$0.750\pm0.009$} and \mbox{$1.280\pm0.013$}, respectively, where the uncertainties are due to the size of the simulated samples.


Most of the systematic effects cancel out in the BF ratio due to the similar topology between signal and normalization channels. The remaining sources of systematic uncertainty of the BF ratio measurement are summarized in Table~\ref{tab:bf_syst}. 
The total systematic uncertainty is determined from the sum in quadrature of all contributions.

The tracking efficiencies of charged pions and kaons mostly cancel out in the ratios of Eq.~\eqref{eq:BFratio}, except for the potential differences of their hadronic interactions with detector materials. This uncertainty per track is estimated to be 1.4\% for pions and 1.1\% for kaons~\cite{LHCb-DP-2013-002}.
Hence, their sum in quadrature, 1.78\%, is assigned as a systematic uncertainty, assuming the uncertainties between pions and kaons are uncorrelated.
The PID variables from the simulated samples are corrected to match the large high-purity calibration samples~\cite{cite:pidgen}. The difference between the total efficiency ratios between the PID transformation method and the PID resampling method~\cite{cite:pidgen} is assigned as systematic uncertainty.
The systematic uncertainty due to the hardware trigger requirement is also studied. 
The trigger efficiency is assumed to vary as a function of the momentum of the \Omegac baryon. Owing to the limited signal yields of both signal channels in data, the trigger efficiency is studied for the normalization channel to understand the difference between data and simulation, which is then used to correct the efficiencies of the signal and normalization modes.
The difference between the corrected efficiency ratio and the uncorrected ratio is assigned as a systematic uncertainty.

The choice of analytical probability density functions to model the fit components affects the determination of the signal yields. Here, the systematic uncertainty is obtained by varying the invariant-mass fit functions of all decay channels following the aforementioned method used in the \Omegac mass measurement.

The simulated samples are generated without considering any asymmetry in the angular distributions for charmed weak decays, given the lack of knowledge of the dynamics of the \Omegac decays. The systematic uncertainty associated with the decay model used in the simulation is evaluated by a simultaneous reweighting of the different angular variables in the simulated samples to the corresponding signal-weighted data distributions~\cite{Pivk:2004ty}.
The uncertainty from the \Omegam lifetime cancels in the ratio $\mathcal B(\Omegam\Kp)/\mathcal B(\Omegam\pip)$. For the ratio $\mathcal B(\Xim\pip)/\mathcal B(\Omegam\pip)$, the \Omegam and \Xim lifetimes are varied within 1 standard deviation of the world averages~\cite{PDG2022}, and the corresponding efficiency ratios are re-estimated. The maximum change, 0.59\%, is taken as the systematic uncertainty.
The uncertainty of the signal efficiency due to the finite simulation sample size is assigned as an additional systematic uncertainty.

To estimate the systematic uncertainty linked to the signal-weighting strategy, the weights applied are extracted from the \OcToXipi decay mode, which has a higher yield among the signal channels, instead of those obtained from the normalization mode.
The efficiency of the signal channel is recalculated, and the change in the ratio is taken as a systematic uncertainty from the weighting strategy.

The ratio of invariant-mass resolutions between data and simulation is assumed to depend linearly on the difference between the mass of \Omegac baryon and the sum of the masses of its decay products. Thus, by performing a linear fit to the ratio of the invariant-mass resolution for the three decay processes, a corrected signal mass resolution can be obtained for each decay. Pseudoexperiments are generated with the baseline model and fitted with the corrected resolution model. 
The difference in signal yields obtained by the baseline and alternative model is taken as the systematic uncertainty due to the mass resolution.

For the \OcToXipi decay process, the external inputs of \mbox{$\mathcal{B}(\OmToLzK)$} and \mbox{$\mathcal{B}(\XiToLzpi)$} are taken from the known values~\cite{PDG2022} and the uncertainties are propagated to the measured BF ratio.

\begin{table}[tp]
\setlength{\abovecaptionskip}{0cm}
\setlength{\belowcaptionskip}{-0.3cm}
  \caption{
    Systematic uncertainties (in percentages) for the BF ratio measurement.}
\begin{center}\begin{tabular}{lcc}
    \hline
    Source & $\mathcal B(\Omegam\Kp)/\mathcal B(\Omegam\pip)$ & $\mathcal B(\Xim\pip)/\mathcal B(\Omegam\pip)$\\
    \hline
    Tracking efficiency & $1.78$ & $1.78$\\
    PID efficiency & $3.37$ & $0.62$\\
    Trigger efficiency & $1.26$ & $0.69$\\
    Fit model & $0.16$ & $0.54$\\
    Decay model & $3.59$ & $1.32$\\
    Lifetimes of \Omegam and \Xim  &  & $0.59$\\
    Simulation sample size & $0.07$ & $0.08$\\
    Reweight strategy & $2.82$ & $0.52$\\
    Mass resolution & $2.35$ & $0.97$ \\
    \hline
    Total & $6.51$ & $2.76$\\
    \hline
    External input BFs &  & $1.04$\\
    \hline
  \end{tabular}\end{center}
\label{tab:bf_syst}
\end{table}


In conclusion, using \proton\proton collision data collected with the \lhcb experiment at a center-of-mass energy of 13\tev, corresponding to an integrated luminosity of \mbox{$5.4\invfb$}, the first observation of the \OcToOmK and \OcToXipi singly Cabibbo-suppressed decays is reported. 
The BF ratios are measured to be
\begin{eqnarray}
    \frac{\mathcal{B}(\OcToOmK)}{\mathcal{B}(\OcToOmpi)}&=&[6.08\pm0.51\stat\pm 0.40\syst]\%\text{,}\nonumber\\
    \frac{\mathcal{B}(\OcToXipi)}{\mathcal{B}(\OcToOmpi)}&=&[15.81\pm0.87\stat\pm0.44\syst\pm0.16\ext]\%\text{,}\nonumber
\end{eqnarray}
where the third uncertainty for the \OcToXipi decay is due to the external BF inputs to the measurement.
In addition, the \Omegac mass is measured to be 
\begin{eqnarray}
M(\Omegac)&=&2695.28\pm0.07\stat\pm0.27\syst\pm0.30\ext\mev\text{.}\nonumber
\end{eqnarray}
This is the most precise measurement of the \Omegac mass to date, and improves the precision of the present world average~\cite{PDG2022} by a factor of 4.
This \Omegac mass measurement provides a strict constraint on various theoretical models.
The mass difference with respect to the \Omegam mass is found to be
\begin{eqnarray}
M(\Omegac)-M(\Omegam)&=&1022.83\pm0.07\stat\pm0.27\syst\mev\text{.}\nonumber
\end{eqnarray}

The BF ratio \mbox{$\mathcal{B}(\OcToXipi)/\mathcal{B}(\OcToOmpi)$} reported in this Letter is larger than the estimated value of 10.38\% from the current algebra calculation with factorizable and nonfactorizable amplitudes~\cite{cite:predict_xu}, while it is further away from the predicted value of 3.45\% from the light-front quark model using only the external \W-emission contribution~\cite{cite:predict_tsai,cite:predict_zhao}.
Additionally, assuming negligible nonfactorizable contributions and a relevant form factor similar to that of \OcToOmpi, the BF ratio \mbox{$\mathcal{B}(\OcToOmK)/\mathcal{B}(\OcToOmpi)$} can be estimated to be \mbox{$(|V_{us}|^{2}/|V_{ud}|^{2})\times R_{\rm phsp}\approx0.0467\pm0.0003$}~\cite{PDG2022}, where \mbox{$|V_{us}|$} and \mbox{$|V_{ud}|$} are CKM matrix elements, and \mbox{$R_{\rm phsp}$} is the ratio of phase-space factors.
This predicted value is more than \mbox{$2\,\sigma$} smaller than the measurement presented in this Letter.
These results indicate that the nonfactorizable contributions are necessary to accurately calculate the BFs in both \OcToOmK and \OcToXipi decays, and provide unique and fresh inputs to understand the nonperturbative effects in models based on quantum chromodynamics.

\section*{Acknowledgements}
%
%
\noindent We express our gratitude to our colleagues in the CERN
accelerator departments for the excellent performance of the LHC. We
thank the technical and administrative staff at the LHCb
institutes.
We acknowledge support from CERN and from the national agencies:
CAPES, CNPq, FAPERJ and FINEP (Brazil); 
MOST and NSFC (China); 
CNRS/IN2P3 (France); 
BMBF, DFG and MPG (Germany); 
INFN (Italy); 
NWO (Netherlands); 
MNiSW and NCN (Poland); 
MEN/IFA (Romania); 
MICINN (Spain); 
SNSF and SER (Switzerland); 
NASU (Ukraine); 
STFC (United Kingdom); 
DOE NP and NSF (USA).
We acknowledge the computing resources that are provided by CERN, IN2P3
(France), KIT and DESY (Germany), INFN (Italy), SURF (Netherlands),
PIC (Spain), GridPP (United Kingdom), 
CSCS (Switzerland), IFIN-HH (Romania), CBPF (Brazil),
Polish WLCG  (Poland), and NERSC (USA).
We are indebted to the communities behind the multiple open-source
software packages on which we depend.
Individual groups or members have received support from
ARC and ARDC (Australia);
Minciencias (Colombia);
AvH Foundation (Germany);
EPLANET, Marie Sk\l{}odowska-Curie Actions, ERC and NextGenerationEU (European Union);
A*MIDEX, ANR, IPhU and Labex P2IO, and R\'{e}gion Auvergne-Rh\^{o}ne-Alpes (France);
Key Research Program of Frontier Sciences of CAS, CAS PIFI, CAS CCEPP, 
Fundamental Research Funds for the Central Universities, 
and Sci. \& Tech. Program of Guangzhou (China);
GVA, XuntaGal, GENCAT, Inditex, InTalent and Prog.~Atracci\'on Talento, CM (Spain);
SRC (Sweden);
and the Leverhulme Trust, the Royal Society,
 and UKRI (United Kingdom).

\addcontentsline{toc}{section}{References}
\setboolean{inbibliography}{true}
\bibliographystyle{LHCb}
\bibliography{main,standard,LHCb-PAPER,LHCb-CONF,LHCb-DP,LHCb-TDR}

\ifx\mcitethebibliography\mciteundefinedmacro
\PackageError{LHCb.bst}{mciteplus.sty has not been loaded}
{This bibstyle requires the use of the mciteplus package.}\fi
\providecommand{\href}[2]{#2}
\begin{mcitethebibliography}{10}
\mciteSetBstSublistMode{n}
\mciteSetBstMaxWidthForm{subitem}{\alph{mcitesubitemcount})}
\mciteSetBstSublistLabelBeginEnd{\mcitemaxwidthsubitemform\space}
{\relax}{\relax}

\bibitem{cite:Cheng_CBP}
H.~Cheng, \ifthenelse{\boolean{articletitles}}{\emph{{Charmed baryon physics circa 2021}}, }{}\href{https://doi.org/10.1016/j.cjph.2022.06.021}{Chin.\ J.\ Phys.\  \textbf{78} (2022) 324}, \href{http://arxiv.org/abs/2109.01216}{{\normalfont\ttfamily arXiv:2109.01216}}\relax
\mciteBstWouldAddEndPuncttrue
\mciteSetBstMidEndSepPunct{\mcitedefaultmidpunct}
{\mcitedefaultendpunct}{\mcitedefaultseppunct}\relax
\EndOfBibitem
\bibitem{PDG2022}
Particle Data Group, R.~L. Workman {\em et~al.}, \ifthenelse{\boolean{articletitles}}{\emph{{\href{https://academic.oup.com/ptep/article/2022/8/083C01/6651666}{Review of Particle Physics}}}, }{}\href{https://doi.org/10.1093/ptep/ptac097}{Prog.\ Theor.\ Exp.\ Phys.\  \textbf{2022} (2022) 083C01}\relax
\mciteBstWouldAddEndPuncttrue
\mciteSetBstMidEndSepPunct{\mcitedefaultmidpunct}
{\mcitedefaultendpunct}{\mcitedefaultseppunct}\relax
\EndOfBibitem
\bibitem{cite:omc0_mass_E687}
E687 collaboration, P.~L. Frabetti {\em et~al.}, \ifthenelse{\boolean{articletitles}}{\emph{{Observation and mass measurement of $\Omegac\rightarrow\Sigmap\Km\Km\pip$}}, }{}\href{https://doi.org/10.1016/0370-2693(94)91351-X}{Phys.\ Lett.\  \textbf{B338} (1994) 106}\relax
\mciteBstWouldAddEndPuncttrue
\mciteSetBstMidEndSepPunct{\mcitedefaultmidpunct}
{\mcitedefaultendpunct}{\mcitedefaultseppunct}\relax
\EndOfBibitem
\bibitem{cite:omc0_mass_cleo}
CLEO collaboration, D.~Cronin-Hennessy {\em et~al.}, \ifthenelse{\boolean{articletitles}}{\emph{{Observation of the $\Omegac$ charmed baryon at CLEO}}, }{}\href{https://doi.org/10.1103/PhysRevLett.86.3730}{Phys.\ Rev.\ Lett.\  \textbf{86} (2001) 3730}\relax
\mciteBstWouldAddEndPuncttrue
\mciteSetBstMidEndSepPunct{\mcitedefaultmidpunct}
{\mcitedefaultendpunct}{\mcitedefaultseppunct}\relax
\EndOfBibitem
\bibitem{cite:omc0_mass_belle}
Belle collaboration, E.~Solovieva {\em et~al.}, \ifthenelse{\boolean{articletitles}}{\emph{{Study of the $\Omegac$ and $\it{\Omega^{*0}_c}$ baryons at Belle}}, }{}\href{https://doi.org/10.1016/j.physletb.2008.12.062}{Phys.\ Lett.\  \textbf{B672} (2009) 1}\relax
\mciteBstWouldAddEndPuncttrue
\mciteSetBstMidEndSepPunct{\mcitedefaultmidpunct}
{\mcitedefaultendpunct}{\mcitedefaultseppunct}\relax
\EndOfBibitem
\bibitem{LHCb-PAPER-2018-028}
LHCb collaboration, R.~Aaij {\em et~al.}, \ifthenelse{\boolean{articletitles}}{\emph{{Measurement of the $\Omegares_c^0$ lifetime}}, }{}\href{https://doi.org/10.1103/PhysRevLett.121.092003}{Phys.\ Rev.\ Lett.\  \textbf{121} (2018) 092003}, \href{http://arxiv.org/abs/1807.02024}{{\normalfont\ttfamily arXiv:1807.02024}}\relax
\mciteBstWouldAddEndPuncttrue
\mciteSetBstMidEndSepPunct{\mcitedefaultmidpunct}
{\mcitedefaultendpunct}{\mcitedefaultseppunct}\relax
\EndOfBibitem
\bibitem{LHCb-PAPER-2021-021}
LHCb collaboration, R.~Aaij {\em et~al.}, \ifthenelse{\boolean{articletitles}}{\emph{{Measurement of the lifetimes of promptly produced $\Omegac$ and $\Xicz$ baryons}}, }{}\href{https://doi.org/10.1016/j.scib.2021.11.022}{Science Bulletin \textbf{67} (2022) 479}, \href{http://arxiv.org/abs/2109.01334}{{\normalfont\ttfamily arXiv:2109.01334}}\relax
\mciteBstWouldAddEndPuncttrue
\mciteSetBstMidEndSepPunct{\mcitedefaultmidpunct}
{\mcitedefaultendpunct}{\mcitedefaultseppunct}\relax
\EndOfBibitem
\bibitem{Belle-II:2022plj}
Belle II collaboration, F.~J. Abudinen {\em et~al.}, \ifthenelse{\boolean{articletitles}}{\emph{{Measurement of the $\Omegac$ lifetime at Belle II}}, }{}\href{https://doi.org/10.1103/PhysRevD.107.L031103}{Phys.\ Rev.\  \textbf{D107} (2023) L031103}, \href{http://arxiv.org/abs/2208.08573}{{\normalfont\ttfamily arXiv:2208.08573}}\relax
\mciteBstWouldAddEndPuncttrue
\mciteSetBstMidEndSepPunct{\mcitedefaultmidpunct}
{\mcitedefaultendpunct}{\mcitedefaultseppunct}\relax
\EndOfBibitem
\bibitem{Cheng:2021vca}
H.-Y. Cheng, \ifthenelse{\boolean{articletitles}}{\emph{{The strangest lifetime: A bizarre story of $\tau(\Omegac)$}}, }{}\href{https://doi.org/10.1016/j.scib.2021.11.025}{Science Bulletin \textbf{67(5)} (2022) 445}, \href{http://arxiv.org/abs/2111.09566}{{\normalfont\ttfamily arXiv:2111.09566}}\relax
\mciteBstWouldAddEndPuncttrue
\mciteSetBstMidEndSepPunct{\mcitedefaultmidpunct}
{\mcitedefaultendpunct}{\mcitedefaultseppunct}\relax
\EndOfBibitem
\bibitem{cite:pole_model}
H.~Y. Cheng and B.~Tseng, \ifthenelse{\boolean{articletitles}}{\emph{{Cabibbo-allowed nonleptonic weak decays of charmed baryons}}, }{}\href{https://doi.org/10.1103/PhysRevD.48.4188}{Phys.\ Rev.\  \textbf{D48} (1993) 4188}\relax
\mciteBstWouldAddEndPuncttrue
\mciteSetBstMidEndSepPunct{\mcitedefaultmidpunct}
{\mcitedefaultendpunct}{\mcitedefaultseppunct}\relax
\EndOfBibitem
\bibitem{cite:Omegac_Omegapi}
H.~Y. Cheng, \ifthenelse{\boolean{articletitles}}{\emph{{Nonleptonic weak decays of bottom baryons}}, }{}\href{https://doi.org/10.1103/PhysRevD.99.079901}{Phys.\ Rev.\  \textbf{D99} (2019) 079901}, \href{http://arxiv.org/abs/hep-ph/9612223}{{\normalfont\ttfamily arXiv:hep-ph/9612223}}\relax
\mciteBstWouldAddEndPuncttrue
\mciteSetBstMidEndSepPunct{\mcitedefaultmidpunct}
{\mcitedefaultendpunct}{\mcitedefaultseppunct}\relax
\EndOfBibitem
\bibitem{cite:predict_xuq}
Q.~P. Xu and A.~N. Kamal, \ifthenelse{\boolean{articletitles}}{\emph{{Nonleptonic charmed-baryon decays: \mbox{$B_c\rightarrow B(\frac{3}{2}^+, {\rm decuplet})+P(0^-)~{\rm or}~V(1^-)$}}}, }{}\href{https://doi.org/10.1103/PhysRevD.46.3836}{Phys.\ Rev.\  \textbf{D46} (1992) 3836}\relax
\mciteBstWouldAddEndPuncttrue
\mciteSetBstMidEndSepPunct{\mcitedefaultmidpunct}
{\mcitedefaultendpunct}{\mcitedefaultseppunct}\relax
\EndOfBibitem
\bibitem{cite:cov1}
J.~G. K{\"o}rner and M.~Kramer, \ifthenelse{\boolean{articletitles}}{\emph{{Exclusive non-leptonic charm baryon decays}}, }{}\href{https://doi.org/10.1007/BF01561305}{Z.\ Phys.\  \textbf{C55} (1992) 659}\relax
\mciteBstWouldAddEndPuncttrue
\mciteSetBstMidEndSepPunct{\mcitedefaultmidpunct}
{\mcitedefaultendpunct}{\mcitedefaultseppunct}\relax
\EndOfBibitem
\bibitem{cite:cov2}
M.~A. Ivanov, J.~G. K{\"o}rner, V.~E. Lyubovitskij, and A.~G. Rusetsky, \ifthenelse{\boolean{articletitles}}{\emph{{Exclusive nonleptonic decays of bottom and charm baryons in a relativistic three-quark model: Evaluation of nonfactorizing diagrams}}, }{}\href{https://doi.org/10.1103/PhysRevD.57.5632}{Phys.\ Rev.\  \textbf{C57} (1998) 5632}\relax
\mciteBstWouldAddEndPuncttrue
\mciteSetBstMidEndSepPunct{\mcitedefaultmidpunct}
{\mcitedefaultendpunct}{\mcitedefaultseppunct}\relax
\EndOfBibitem
\bibitem{cite:predict_guts}
T.~Gutsche, M.~A. Ivanov, J.~G. K{\"o}rner, and V.~E. Lyubovitskij, \ifthenelse{\boolean{articletitles}}{\emph{{Nonleptonic two-body decays of single heavy baryons $\it{\Lambda_Q, \Xi_Q,}$ and $\it{\Omega_Q (Q=b,c)}$ induced by $W$ emission in the covariant confined quark model}}, }{}\href{https://doi.org/10.1103/PhysRevD.98.074011}{Phys.\ Rev.\  \textbf{D98} (2018) 074011}\relax
\mciteBstWouldAddEndPuncttrue
\mciteSetBstMidEndSepPunct{\mcitedefaultmidpunct}
{\mcitedefaultendpunct}{\mcitedefaultseppunct}\relax
\EndOfBibitem
\bibitem{cite:predict_xu}
S.~Hu, G.~Meng, and F.~Xu, \ifthenelse{\boolean{articletitles}}{\emph{{Hadronic weak decays of the charmed baryon $\Omegac$}}, }{}\href{https://doi.org/10.1103/PhysRevD.101.094033}{Phys.\ Rev.\  \textbf{D101} (2020) 094033}, \href{http://arxiv.org/abs/2003.04705}{{\normalfont\ttfamily arXiv:2003.04705}}\relax
\mciteBstWouldAddEndPuncttrue
\mciteSetBstMidEndSepPunct{\mcitedefaultmidpunct}
{\mcitedefaultendpunct}{\mcitedefaultseppunct}\relax
\EndOfBibitem
\bibitem{cite:predict_tsai}
Y.~K. Hsiao, L.~Yang, C.~C. Lih, and S.~Y. Tsai, \ifthenelse{\boolean{articletitles}}{\emph{{Charmed $\Omegac$ weak decays into $\it{\Omega}$ in the light-front quark model}}, }{}\href{https://doi.org/10.1140/epjc/s10052-020-08619-y}{Eur.\ Phys.\ J.\  \textbf{C80} (2020) 1066}, \href{http://arxiv.org/abs/2009.12752}{{\normalfont\ttfamily arXiv:2009.12752}}\relax
\mciteBstWouldAddEndPuncttrue
\mciteSetBstMidEndSepPunct{\mcitedefaultmidpunct}
{\mcitedefaultendpunct}{\mcitedefaultseppunct}\relax
\EndOfBibitem
\bibitem{cite:predict_zhao}
Z.~X. Zhao, \ifthenelse{\boolean{articletitles}}{\emph{{Weak decays of heavy baryons in the light-front approach}}, }{}\href{https://doi.org/10.1088/1674-1137/42/9/093101}{Chin.\ Phys.\  \textbf{C42} (2018) 093101}\relax
\mciteBstWouldAddEndPuncttrue
\mciteSetBstMidEndSepPunct{\mcitedefaultmidpunct}
{\mcitedefaultendpunct}{\mcitedefaultseppunct}\relax
\EndOfBibitem
\bibitem{cite:belle_xipi_omk}
Belle collaboration, S.~Jia {\em et~al.}, \ifthenelse{\boolean{articletitles}}{\emph{{Evidence for the singly Cabibbo-suppressed decay $\Omegac\rightarrow\Xim\pip$ and search for $\Omegac\rightarrow\Xim\Kp$ and $\Omegam\Kp$ decays at Belle}}, }{}\href{https://doi.org/10.1007/JHEP01(2023)055}{JHEP \textbf{01} (2023) 055}, \href{http://arxiv.org/abs/2209.08583}{{\normalfont\ttfamily arXiv:2209.08583}}\relax
\mciteBstWouldAddEndPuncttrue
\mciteSetBstMidEndSepPunct{\mcitedefaultmidpunct}
{\mcitedefaultendpunct}{\mcitedefaultseppunct}\relax
\EndOfBibitem
\bibitem{cite:LQCD1}
C.~Alexandrou {\em et~al.}, \ifthenelse{\boolean{articletitles}}{\emph{{Low-lying baryon masses using twisted mass fermions ensembles at the physical pion mass}}, }{}\href{http://arxiv.org/abs/2309.04401}{{\normalfont\ttfamily arXiv:2309.04401}}\relax
\mciteBstWouldAddEndPuncttrue
\mciteSetBstMidEndSepPunct{\mcitedefaultmidpunct}
{\mcitedefaultendpunct}{\mcitedefaultseppunct}\relax
\EndOfBibitem
\bibitem{cite:LQCD2}
P.~P\'{e}rez-Rubio, S.~Collins, and G.~S. Bali, \ifthenelse{\boolean{articletitles}}{\emph{{Charmed baryon spectroscopy and light flavor symmetry from lattice QCD}}, }{}\href{https://doi.org/10.1103/PhysRevD.92.034504}{Phys.\ Rev.\  \textbf{D92} (2015) 034504}, \href{http://arxiv.org/abs/1503.08440}{{\normalfont\ttfamily arXiv:1503.08440}}\relax
\mciteBstWouldAddEndPuncttrue
\mciteSetBstMidEndSepPunct{\mcitedefaultmidpunct}
{\mcitedefaultendpunct}{\mcitedefaultseppunct}\relax
\EndOfBibitem
\bibitem{cite:LQCD3}
R.~A. Briceno, H.-W. Lin, and D.~R. Bolton, \ifthenelse{\boolean{articletitles}}{\emph{{Charmed-baryon spectroscopy from lattice QCD with $N_f=2+1+1$ flavors}}, }{}\href{https://doi.org/10.1103/PhysRevD.86.094504}{Phys.\ Rev.\  \textbf{D86} (2012) 094504}, \href{http://arxiv.org/abs/1207.3536}{{\normalfont\ttfamily arXiv:1207.3536}}\relax
\mciteBstWouldAddEndPuncttrue
\mciteSetBstMidEndSepPunct{\mcitedefaultmidpunct}
{\mcitedefaultendpunct}{\mcitedefaultseppunct}\relax
\EndOfBibitem
\bibitem{cite:LQCD4}
Z.~S. Brown {\em et~al.}, \ifthenelse{\boolean{articletitles}}{\emph{{Charmed bottom baryon spectroscopy from lattice QCD}}, }{}\href{https://doi.org/10.1103/PhysRevD.90.094507}{Phys.\ Rev.\  \textbf{D90} (2014) 094507}, \href{http://arxiv.org/abs/1409.0497}{{\normalfont\ttfamily arXiv:1409.0497}}\relax
\mciteBstWouldAddEndPuncttrue
\mciteSetBstMidEndSepPunct{\mcitedefaultmidpunct}
{\mcitedefaultendpunct}{\mcitedefaultseppunct}\relax
\EndOfBibitem
\bibitem{cite:LQCD5}
L.~Liu {\em et~al.}, \ifthenelse{\boolean{articletitles}}{\emph{{Singly and doubly charmed $J=1/2$ baryon spectrum from lattice QCD}}, }{}\href{https://doi.org/10.1103/PhysRevD.81.094505}{Phys.\ Rev.\  \textbf{D81} (2010) 094505}, \href{http://arxiv.org/abs/0909.3294}{{\normalfont\ttfamily arXiv:0909.3294}}\relax
\mciteBstWouldAddEndPuncttrue
\mciteSetBstMidEndSepPunct{\mcitedefaultmidpunct}
{\mcitedefaultendpunct}{\mcitedefaultseppunct}\relax
\EndOfBibitem
\bibitem{cite:LQCD6}
S.~Durr, G.~Koutsou, and T.~Lippert, \ifthenelse{\boolean{articletitles}}{\emph{{Meson and baryon dispersion relations with Brillouin fermions}}, }{}\href{https://doi.org/10.1103/PhysRevD.86.114514}{Phys.\ Rev.\  \textbf{D86} (2012) 114514}, \href{http://arxiv.org/abs/1208.6270}{{\normalfont\ttfamily arXiv:1208.6270}}\relax
\mciteBstWouldAddEndPuncttrue
\mciteSetBstMidEndSepPunct{\mcitedefaultmidpunct}
{\mcitedefaultendpunct}{\mcitedefaultseppunct}\relax
\EndOfBibitem
\bibitem{cite:LQCD7}
PACS-CS collaboration, Y.~Namekawa {\em et~al.}, \ifthenelse{\boolean{articletitles}}{\emph{{Charmed baryons at the physical point in $2+1$ flavor lattice QCD}}, }{}\href{https://doi.org/10.1103/PhysRevD.87.094512}{Phys.\ Rev.\  \textbf{D87} (2013) 094512}, \href{http://arxiv.org/abs/1301.4743}{{\normalfont\ttfamily arXiv:1301.4743}}\relax
\mciteBstWouldAddEndPuncttrue
\mciteSetBstMidEndSepPunct{\mcitedefaultmidpunct}
{\mcitedefaultendpunct}{\mcitedefaultseppunct}\relax
\EndOfBibitem
\bibitem{cite:LQCD8}
S.~Samuel and K.~Moriarty, \ifthenelse{\boolean{articletitles}}{\emph{{Beautiful mass predictions from scalar lattice QCD}}, }{}\href{https://doi.org/10.1016/0370-2693(86)90715-X}{Phys.\ Lett.\  \textbf{B175} (19986) 197}\relax
\mciteBstWouldAddEndPuncttrue
\mciteSetBstMidEndSepPunct{\mcitedefaultmidpunct}
{\mcitedefaultendpunct}{\mcitedefaultseppunct}\relax
\EndOfBibitem
\bibitem{cite:Other1}
E.~Jenkins, \ifthenelse{\boolean{articletitles}}{\emph{{Heavy baryon masses in the $1/m_Q$ and $1/N_c$ expansions}}, }{}\href{https://doi.org/10.1103/PhysRevD.54.4515}{Phys.\ Rev.\  \textbf{D54} (1996) 4515}\relax
\mciteBstWouldAddEndPuncttrue
\mciteSetBstMidEndSepPunct{\mcitedefaultmidpunct}
{\mcitedefaultendpunct}{\mcitedefaultseppunct}\relax
\EndOfBibitem
\bibitem{cite:Other2}
R.~Roncaglia {\em et~al.}, \ifthenelse{\boolean{articletitles}}{\emph{{Predicting the masses of baryons containing one or two heavy quarks}}, }{}\href{https://doi.org/10.1103/PhysRevD.52.1722}{Phys.\ Rev.\  \textbf{D52} (1995) 1722}\relax
\mciteBstWouldAddEndPuncttrue
\mciteSetBstMidEndSepPunct{\mcitedefaultmidpunct}
{\mcitedefaultendpunct}{\mcitedefaultseppunct}\relax
\EndOfBibitem
\bibitem{cite:Other3}
A.~Martin and J.~Richard, \ifthenelse{\boolean{articletitles}}{\emph{{$\it \Omega_c$ and other heavy baryons revisited}}, }{}\href{https://doi.org/10.1016/0370-2693(87)91029-X}{Phys.\ Lett.\  \textbf{B355} (1995) 345}\relax
\mciteBstWouldAddEndPuncttrue
\mciteSetBstMidEndSepPunct{\mcitedefaultmidpunct}
{\mcitedefaultendpunct}{\mcitedefaultseppunct}\relax
\EndOfBibitem
\bibitem{cite:Other4}
M.~Rho {\em et~al.}, \ifthenelse{\boolean{articletitles}}{\emph{{Charmed baryons as soliton-$D$ meson bound states}}, }{}\href{https://doi.org/10.1016/0370-2693(90)90802-D}{Phys.\ Lett.\  \textbf{B251} (1990) 597}\relax
\mciteBstWouldAddEndPuncttrue
\mciteSetBstMidEndSepPunct{\mcitedefaultmidpunct}
{\mcitedefaultendpunct}{\mcitedefaultseppunct}\relax
\EndOfBibitem
\bibitem{cite:Other5}
D.~Izatt {\em et~al.}, \ifthenelse{\boolean{articletitles}}{\emph{{Spectroscopy of hadrons containing one heavy quark}}, }{}\href{https://doi.org/10.1016/0550-3213(82)90347-9}{Nucl.\ Phys.\  \textbf{B199} (1982) 269}\relax
\mciteBstWouldAddEndPuncttrue
\mciteSetBstMidEndSepPunct{\mcitedefaultmidpunct}
{\mcitedefaultendpunct}{\mcitedefaultseppunct}\relax
\EndOfBibitem
\bibitem{LHCb-DP-2008-001}
LHCb collaboration, A.~A. Alves~Jr.\ {\em et~al.}, \ifthenelse{\boolean{articletitles}}{\emph{{The \lhcb detector at the LHC}}, }{}\href{https://doi.org/10.1088/1748-0221/3/08/S08005}{JINST \textbf{3} (2008) S08005}\relax
\mciteBstWouldAddEndPuncttrue
\mciteSetBstMidEndSepPunct{\mcitedefaultmidpunct}
{\mcitedefaultendpunct}{\mcitedefaultseppunct}\relax
\EndOfBibitem
\bibitem{LHCb-DP-2014-002}
LHCb collaboration, R.~Aaij {\em et~al.}, \ifthenelse{\boolean{articletitles}}{\emph{{LHCb detector performance}}, }{}\href{https://doi.org/10.1142/S0217751X15300227}{Int.\ J.\ Mod.\ Phys.\  \textbf{A30} (2015) 1530022}, \href{http://arxiv.org/abs/1412.6352}{{\normalfont\ttfamily arXiv:1412.6352}}\relax
\mciteBstWouldAddEndPuncttrue
\mciteSetBstMidEndSepPunct{\mcitedefaultmidpunct}
{\mcitedefaultendpunct}{\mcitedefaultseppunct}\relax
\EndOfBibitem
\bibitem{LHCb-DP-2014-001}
R.~Aaij {\em et~al.}, \ifthenelse{\boolean{articletitles}}{\emph{{Performance of the LHCb Vertex Locator}}, }{}\href{https://doi.org/10.1088/1748-0221/9/09/P09007}{JINST \textbf{9} (2014) P09007}, \href{http://arxiv.org/abs/1405.7808}{{\normalfont\ttfamily arXiv:1405.7808}}\relax
\mciteBstWouldAddEndPuncttrue
\mciteSetBstMidEndSepPunct{\mcitedefaultmidpunct}
{\mcitedefaultendpunct}{\mcitedefaultseppunct}\relax
\EndOfBibitem
\bibitem{LHCb-DP-2013-003}
R.~Arink {\em et~al.}, \ifthenelse{\boolean{articletitles}}{\emph{{Performance of the LHCb Outer Tracker}}, }{}\href{https://doi.org/10.1088/1748-0221/9/01/P01002}{JINST \textbf{9} (2014) P01002}, \href{http://arxiv.org/abs/1311.3893}{{\normalfont\ttfamily arXiv:1311.3893}}\relax
\mciteBstWouldAddEndPuncttrue
\mciteSetBstMidEndSepPunct{\mcitedefaultmidpunct}
{\mcitedefaultendpunct}{\mcitedefaultseppunct}\relax
\EndOfBibitem
\bibitem{LHCb-DP-2012-002}
A.~A. Alves~Jr.\ {\em et~al.}, \ifthenelse{\boolean{articletitles}}{\emph{{Performance of the LHCb muon system}}, }{}\href{https://doi.org/10.1088/1748-0221/8/02/P02022}{JINST \textbf{8} (2013) P02022}, \href{http://arxiv.org/abs/1211.1346}{{\normalfont\ttfamily arXiv:1211.1346}}\relax
\mciteBstWouldAddEndPuncttrue
\mciteSetBstMidEndSepPunct{\mcitedefaultmidpunct}
{\mcitedefaultendpunct}{\mcitedefaultseppunct}\relax
\EndOfBibitem
\bibitem{LHCb-DP-2019-002}
R.~Aaij {\em et~al.}, \ifthenelse{\boolean{articletitles}}{\emph{{A comprehensive real-time analysis model at the LHCb experiment}}, }{}\href{https://doi.org/10.1088/1748-0221/14/04/P04006}{JINST \textbf{14} (2019) P04006}, \href{http://arxiv.org/abs/1903.01360}{{\normalfont\ttfamily arXiv:1903.01360}}\relax
\mciteBstWouldAddEndPuncttrue
\mciteSetBstMidEndSepPunct{\mcitedefaultmidpunct}
{\mcitedefaultendpunct}{\mcitedefaultseppunct}\relax
\EndOfBibitem
\bibitem{Sjostrand:2007gs}
T.~Sj\"{o}strand, S.~Mrenna, and P.~Skands, \ifthenelse{\boolean{articletitles}}{\emph{{A brief introduction to PYTHIA 8.1}}, }{}\href{https://doi.org/10.1016/j.cpc.2008.01.036}{Comput.\ Phys.\ Commun.\  \textbf{178} (2008) 852}, \href{http://arxiv.org/abs/0710.3820}{{\normalfont\ttfamily arXiv:0710.3820}}\relax
\mciteBstWouldAddEndPuncttrue
\mciteSetBstMidEndSepPunct{\mcitedefaultmidpunct}
{\mcitedefaultendpunct}{\mcitedefaultseppunct}\relax
\EndOfBibitem
\bibitem{LHCb-PROC-2010-056}
I.~Belyaev {\em et~al.}, \ifthenelse{\boolean{articletitles}}{\emph{{Handling of the generation of primary events in Gauss, the LHCb simulation framework}}, }{}\href{https://doi.org/10.1088/1742-6596/331/3/032047}{J.\ Phys.\ Conf.\ Ser.\  \textbf{331} (2011) 032047}\relax
\mciteBstWouldAddEndPuncttrue
\mciteSetBstMidEndSepPunct{\mcitedefaultmidpunct}
{\mcitedefaultendpunct}{\mcitedefaultseppunct}\relax
\EndOfBibitem
\bibitem{Lange:2001uf}
D.~J. Lange, \ifthenelse{\boolean{articletitles}}{\emph{{The EvtGen particle decay simulation package}}, }{}\href{https://doi.org/10.1016/S0168-9002(01)00089-4}{Nucl.\ Instrum.\ Meth.\  \textbf{A462} (2001) 152}\relax
\mciteBstWouldAddEndPuncttrue
\mciteSetBstMidEndSepPunct{\mcitedefaultmidpunct}
{\mcitedefaultendpunct}{\mcitedefaultseppunct}\relax
\EndOfBibitem
\bibitem{davidson2015photos}
N.~Davidson, T.~Przedzinski, and Z.~Was, \ifthenelse{\boolean{articletitles}}{\emph{{PHOTOS interface in C++: Technical and physics documentation}}, }{}\href{https://doi.org/https://doi.org/10.1016/j.cpc.2015.09.013}{Comp.\ Phys.\ Comm.\  \textbf{199} (2016) 86}, \href{http://arxiv.org/abs/1011.0937}{{\normalfont\ttfamily arXiv:1011.0937}}\relax
\mciteBstWouldAddEndPuncttrue
\mciteSetBstMidEndSepPunct{\mcitedefaultmidpunct}
{\mcitedefaultendpunct}{\mcitedefaultseppunct}\relax
\EndOfBibitem
\bibitem{Allison:2006ve}
Geant4 collaboration, J.~Allison {\em et~al.}, \ifthenelse{\boolean{articletitles}}{\emph{{Geant4 developments and applications}}, }{}\href{https://doi.org/10.1109/TNS.2006.869826}{IEEE Trans.\ Nucl.\ Sci.\  \textbf{53} (2006) 270}\relax
\mciteBstWouldAddEndPuncttrue
\mciteSetBstMidEndSepPunct{\mcitedefaultmidpunct}
{\mcitedefaultendpunct}{\mcitedefaultseppunct}\relax
\EndOfBibitem
\bibitem{Agostinelli:2002hh}
Geant4 collaboration, S.~Agostinelli {\em et~al.}, \ifthenelse{\boolean{articletitles}}{\emph{{Geant4: A simulation toolkit}}, }{}\href{https://doi.org/10.1016/S0168-9002(03)01368-8}{Nucl.\ Instrum.\ Meth.\  \textbf{A506} (2003) 250}\relax
\mciteBstWouldAddEndPuncttrue
\mciteSetBstMidEndSepPunct{\mcitedefaultmidpunct}
{\mcitedefaultendpunct}{\mcitedefaultseppunct}\relax
\EndOfBibitem
\bibitem{LHCb-PROC-2011-006}
M.~Clemencic {\em et~al.}, \ifthenelse{\boolean{articletitles}}{\emph{{The \lhcb simulation application, Gauss: Design, evolution and experience}}, }{}\href{https://doi.org/10.1088/1742-6596/331/3/032023}{J.\ Phys.\ Conf.\ Ser.\  \textbf{331} (2011) 032023}\relax
\mciteBstWouldAddEndPuncttrue
\mciteSetBstMidEndSepPunct{\mcitedefaultmidpunct}
{\mcitedefaultendpunct}{\mcitedefaultseppunct}\relax
\EndOfBibitem
\bibitem{LHCb-DP-2018-001}
R.~Aaij {\em et~al.}, \ifthenelse{\boolean{articletitles}}{\emph{{Selection and processing of calibration samples to measure the particle identification performance of the LHCb experiment in Run 2}}, }{}\href{https://doi.org/10.1140/epjti/s40485-019-0050-z}{Eur.\ Phys.\ J.\ Tech.\ Instr.\  \textbf{6} (2018) 1}, \href{http://arxiv.org/abs/1803.00824}{{\normalfont\ttfamily arXiv:1803.00824}}\relax
\mciteBstWouldAddEndPuncttrue
\mciteSetBstMidEndSepPunct{\mcitedefaultmidpunct}
{\mcitedefaultendpunct}{\mcitedefaultseppunct}\relax
\EndOfBibitem
\bibitem{Hulsbergen:2005pu}
W.~D. Hulsbergen, \ifthenelse{\boolean{articletitles}}{\emph{{Decay chain fitting with a Kalman filter}}, }{}\href{https://doi.org/10.1016/j.nima.2005.06.078}{Nucl.\ Instrum.\ Meth.\  \textbf{A552} (2005) 566}, \href{http://arxiv.org/abs/physics/0503191}{{\normalfont\ttfamily arXiv:physics/0503191}}\relax
\mciteBstWouldAddEndPuncttrue
\mciteSetBstMidEndSepPunct{\mcitedefaultmidpunct}
{\mcitedefaultendpunct}{\mcitedefaultseppunct}\relax
\EndOfBibitem
\bibitem{cite:johnson}
N.~L. Johnson, \ifthenelse{\boolean{articletitles}}{\emph{{Systems of frequency curves generated by methods of translation}}, }{}\href{https://doi.org/10.1093/biomet/36.1-2.149}{Biometrika \textbf{36} (1949) 149}\relax
\mciteBstWouldAddEndPuncttrue
\mciteSetBstMidEndSepPunct{\mcitedefaultmidpunct}
{\mcitedefaultendpunct}{\mcitedefaultseppunct}\relax
\EndOfBibitem
\bibitem{LHCb-PAPER-2012-048}
LHCb collaboration, R.~Aaij {\em et~al.}, \ifthenelse{\boolean{articletitles}}{\emph{{Measurements of the \Lb, \Xibm, and \Omegab baryon masses}}, }{}\href{https://doi.org/10.1103/PhysRevLett.110.182001}{Phys.\ Rev.\ Lett.\  \textbf{110} (2013) 182001}, \href{http://arxiv.org/abs/1302.1072}{{\normalfont\ttfamily arXiv:1302.1072}}\relax
\mciteBstWouldAddEndPuncttrue
\mciteSetBstMidEndSepPunct{\mcitedefaultmidpunct}
{\mcitedefaultendpunct}{\mcitedefaultseppunct}\relax
\EndOfBibitem
\bibitem{LHCb-PAPER-2013-011}
LHCb collaboration, R.~Aaij {\em et~al.}, \ifthenelse{\boolean{articletitles}}{\emph{{Precision measurement of \D meson mass differences}}, }{}\href{https://doi.org/10.1007/JHEP06(2013)065}{JHEP \textbf{06} (2013) 065}, \href{http://arxiv.org/abs/1304.6865}{{\normalfont\ttfamily arXiv:1304.6865}}\relax
\mciteBstWouldAddEndPuncttrue
\mciteSetBstMidEndSepPunct{\mcitedefaultmidpunct}
{\mcitedefaultendpunct}{\mcitedefaultseppunct}\relax
\EndOfBibitem
\bibitem{LHCb-PAPER-2010-001}
LHCb collaboration, R.~Aaij {\em et~al.}, \ifthenelse{\boolean{articletitles}}{\emph{{Prompt \KS production in \proton\proton collisions at $\sqs=$0.9\tev}}, }{}\href{https://doi.org/10.1016/j.physletb.2010.08.055}{Phys.\ Lett.\  \textbf{B693} (2010) 69}, \href{http://arxiv.org/abs/1008.3105}{{\normalfont\ttfamily arXiv:1008.3105}}\relax
\mciteBstWouldAddEndPuncttrue
\mciteSetBstMidEndSepPunct{\mcitedefaultmidpunct}
{\mcitedefaultendpunct}{\mcitedefaultseppunct}\relax
\EndOfBibitem
\bibitem{Skwarnicki:1986xj}
T.~Skwarnicki, {\em {A study of the radiative cascade transitions between the Upsilon-prime and Upsilon resonances}}, PhD thesis, Institute of Nuclear Physics, Krakow, 1986, {\href{http://inspirehep.net/record/230779/}{DESY-F31-86-02}}\relax
\mciteBstWouldAddEndPuncttrue
\mciteSetBstMidEndSepPunct{\mcitedefaultmidpunct}
{\mcitedefaultendpunct}{\mcitedefaultseppunct}\relax
\EndOfBibitem
\bibitem{LHCb-DP-2013-002}
LHCb collaboration, R.~Aaij {\em et~al.}, \ifthenelse{\boolean{articletitles}}{\emph{{Measurement of the track reconstruction efficiency at LHCb}}, }{}\href{https://doi.org/10.1088/1748-0221/10/02/P02007}{JINST \textbf{10} (2015) P02007}, \href{http://arxiv.org/abs/1408.1251}{{\normalfont\ttfamily arXiv:1408.1251}}\relax
\mciteBstWouldAddEndPuncttrue
\mciteSetBstMidEndSepPunct{\mcitedefaultmidpunct}
{\mcitedefaultendpunct}{\mcitedefaultseppunct}\relax
\EndOfBibitem
\bibitem{cite:pidgen}
A.~Poluektov, \ifthenelse{\boolean{articletitles}}{\emph{{Kernel density estimation of a multidimensional efficiency profile}}, }{}\href{https://doi.org/10.1088/1748-0221/10/02/P02011}{JINST \textbf{10} (2015) P02011}, \href{http://arxiv.org/abs/1411.5528}{{\normalfont\ttfamily arXiv:1411.5528}}\relax
\mciteBstWouldAddEndPuncttrue
\mciteSetBstMidEndSepPunct{\mcitedefaultmidpunct}
{\mcitedefaultendpunct}{\mcitedefaultseppunct}\relax
\EndOfBibitem
\bibitem{Pivk:2004ty}
M.~Pivk and F.~R. Le~Diberder, \ifthenelse{\boolean{articletitles}}{\emph{{sPlot: A statistical tool to unfold data distributions}}, }{}\href{https://doi.org/10.1016/j.nima.2005.08.106}{Nucl.\ Instrum.\ Meth.\  \textbf{A555} (2005) 356}, \href{http://arxiv.org/abs/physics/0402083}{{\normalfont\ttfamily arXiv:physics/0402083}}\relax
\mciteBstWouldAddEndPuncttrue
\mciteSetBstMidEndSepPunct{\mcitedefaultmidpunct}
{\mcitedefaultendpunct}{\mcitedefaultseppunct}\relax
\EndOfBibitem
\end{mcitethebibliography}
 
\newpage
\centerline
{\large\bf LHCb collaboration}
\begin
{flushleft}
\small
R.~Aaij$^{32}$\lhcborcid{0000-0003-0533-1952},
A.S.W.~Abdelmotteleb$^{51}$\lhcborcid{0000-0001-7905-0542},
C.~Abellan~Beteta$^{45}$,
F.~Abudin{\'e}n$^{51}$\lhcborcid{0000-0002-6737-3528},
T.~Ackernley$^{55}$\lhcborcid{0000-0002-5951-3498},
B.~Adeva$^{41}$\lhcborcid{0000-0001-9756-3712},
M.~Adinolfi$^{49}$\lhcborcid{0000-0002-1326-1264},
P.~Adlarson$^{77}$\lhcborcid{0000-0001-6280-3851},
H.~Afsharnia$^{9}$,
C.~Agapopoulou$^{43}$\lhcborcid{0000-0002-2368-0147}
% ...

{\footnotesize \it

% ...
$^{9}$Universit{\'e} Clermont Auvergne, CNRS/IN2P3, LPC, Clermont-Ferrand, France\\
% ...
$^{32}$Nikhef National Institute for Subatomic Physics, Amsterdam, Netherlands\\
% ...
$^{41}$Instituto Galego de F{\'\i}sica de Altas Enerx{\'\i}as (IGFAE), Universidade de Santiago de Compostela, Santiago de Compostela, Spain\\
% ...
$^{43}$European Organization for Nuclear Research (CERN), Geneva, Switzerland\\
% ...
$^{45}$Physik-Institut, Universit{\"a}t Z{\"u}rich, Z{\"u}rich, Switzerland\\
% ...
$^{49}$H.H. Wills Physics Laboratory, University of Bristol, Bristol, United Kingdom\\
% ...
$^{51}$Department of Physics, University of Warwick, Coventry, United Kingdom\\
% ...
$^{54}$School of Physics and Astronomy, University of Glasgow, Glasgow, United Kingdom\\
$^{55}$Oliver Lodge Laboratory, University of Liverpool, Liverpool, United Kingdom\\
% ...
$^{77}$Department of Physics and Astronomy, Uppsala University, Uppsala, Sweden, associated to $^{54}$\\
% ...
}
\end{flushleft}

\end{document}